\documentclass[twocolumn,showpacs]{revtex4}
\usepackage{graphicx}
\usepackage{dcolumn}

\begin{document}


\title{Phase transition in exotic nuclei along the $N=Z$ line}

\author{M. Hasegawa,$^{1}$ K. Kaneko,$^{2}$
        T. Mizusaki,$^{3}$ and Y. Sun$^{4}$}

\affiliation{
$^{1}${Laboratory of Physics, Fukuoka Dental College,
 Fukuoka 814-0193, Japan} \\
$^{2}${Department of Physics, Kyushu Sangyo University,
 Fukuoka 813-8503, Japan} \\
$^{3}${Institute of Natural Sciences,
 Senshu University, Tokyo 101-8425, Japan } \\
$^{4}${Department of Physics and Joint Institute
 for Nuclear Astrophysics,
 University of Notre Dame, Notre Dame, IN 46556, USA}
}

\date{\today}

\begin{abstract}

The abrupt structure change from the nuclei of $N=Z \le 35$ to
those of $N=Z \ge 36$ is investigated by means of shell model
calculations. The basic features of the even-even and odd-odd
nuclei under consideration are nicely reproduced. A sudden jump of
nucleons into the upper $g_{9/2}d_{5/2}$ shell at $N=Z=36$ is
found to be the main reason that causes the qualitative structure
difference. It is argued that the structure change can be viewed
as a decisive change of the mean field, or a phase transition,
along the $N=Z$ line.

\end{abstract}

\pacs{21.10.-k, 21.10.Re, 21.60.Cs}

\maketitle


Phase transition in a many-body system refers to an abrupt,
qualitative change in wave function or mean field. This subject is
of common interest for many subfields. Nuclei, being finite
quantum systems composed of strongly correlated protons and
neutrons, uniquely display transitional features. Often discussed
in nuclei are two types of phase transition: the phase transition
to superfluidity and to deformation. The first one is easy to
trace with the standard BCS theory which gives the critical
condition \cite{Ring}. The second one, the transition to
deformation, takes place as some control parameters vary along an
isotopic (isotonic) chain, which, for example, brings the system
from a spherical to a deformed region.

The second type of phase transition, also called shape phase
transition \cite{Iachello98}, has been theoretically studied but
mainly by means of algebraic models \cite{Frank06}. Well-suited
cases are those described by an algebraic Hamiltonian with a
dynamical group, where a transition-driving control parameter
appears explicitly in the Hamiltonian. Large-scale shell model
descriptions \cite{Caurier95,Kaneko02} for deformation in
medium-mass nuclei have become possible in recent years. The shell
model can provide a more fundamental basis to the study of shape
phase transition. In fact, the Monte Carlo Shell Model calculation
\cite{Shimizu} has shown the first example. The advantage of a
shell model study is that one may see microscopically the origin
of a phase transition by analyzing the wave functions if
transition-driving mechanism is contained in the effective
interaction. The aim of the present Letter is to carry out such a
study using the spherical shell model, to show a shape phase
transition in $A\sim$ 70 nuclei along the $N=Z$ line, and to
understand the cause of the transition by studying the occupations
in the nucleon orbits.

\begin{figure}[b]
\includegraphics[width=6cm,height=6cm]{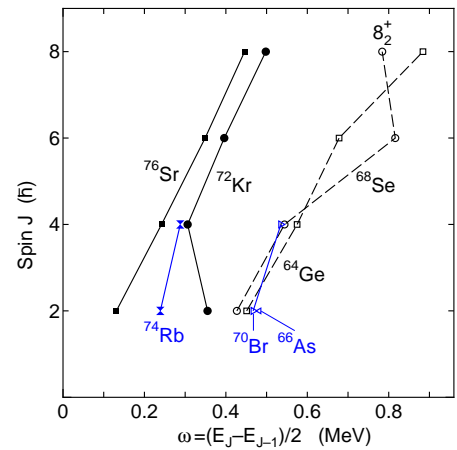}
  \caption{(Color online) The known experimental data shown in the
           graph of spin $J$ versus angular frequency $\omega$.}
  \label{fig1}
\end{figure}

The $N=Z$ nuclei around $A\sim$ 70 exhibit several unique
phenomena \cite{Nazarewicz,Lister90}, and therefore have attracted
many theoretical and experimental studies. This is possibly due to
the rather strong proton-neutron correlations in these nuclei
since protons and neutrons occupy the same orbits. In particular,
there is apparently an abrupt change in structure when the proton
and neutron numbers cross $N=Z=35$, which can be clearly seen from
the graph of spin $J$ versus angular frequency $\omega
=(E(J)-E(J-2))/2$ (the so-called $J-\omega$ graph). Fig.
\ref{fig1} shows a separation of two groups for the $N=Z$ nuclei
from $^{64}$Ge to $^{76}$Sr, one with larger $\omega$'s
corresponding to smaller moments of inertia and another with
smaller $\omega$'s corresponding to larger moments of inertia. In
the left group, $^{76}$Sr takes a straight line whose extension
intersects the origin of the $J-\omega$ graph. Thus $^{76}$Sr
behaves like a rotor with a large and approximately constant
moment of inertia $J/\omega$. The $J-\omega$ curves for $^{72}$Kr
and $^{74}$Rb show that these nuclei have a structure rather
similar to that of $^{76}$Sr although the lowest state of
$^{72}$Kr appears to be peculiar (see discussions below). In
contrast, the nuclei in the right group with $N=Z\le 35$, which
resemble each other, have a moment of inertia only approximately
half of the values of the left group. This sharp difference
suggests a qualitative structure change between the nuclei with
$N=Z\le 35$ and $N=Z\ge 36$. No nuclei sit in between, implying
that the structure change with nucleon number is sudden.

The above structure change along the $N=Z$ line may be comparable
to that near $N=40$ in Ge isotopes \cite{Lebrun}, which has been
observed in the $(p,t)$ cross sections and $B(E2;0_1^+ \rightarrow
2_1^+)$ values. For example, $^{72}$Ge has an unusually low second
$0^+$ state below the $2_1^+$ state, and so does $^{72}$Kr
\cite{Bouchez}. In a separate study for Ge isotopes \cite{HaseN},
we have discovered the possible sources for the structure change
near $N=40$. The study suggests that the change is caused mainly
by a sudden jump of nucleons into the $g_{9/2}$ orbit. We may
expect that the $g_{9/2}$ occupation is also the leading source
for the structure change along the $N=Z$ line.

In the early study \cite{HaseN}, we found it difficult to get a
sufficient $g_{9/2}$ contribution within the $pf_{5/2}g_{9/2}$
shell model, in spite of the fact that this model space is capable
of describing the nuclei $^{64}$Ge and $^{68}$Se \cite{Kaneko}. On
the other hand, the Shell Model Monte Carlo calculations
\cite{Langanke2} suggest that the $d_{5/2}$ orbit has a
cooperative effect that enhances the $g_{9/2}$ contribution. Thus
the $pf_{5/2}g_{9/2}d_{5/2}$ shell model can be a hopeful model to
investigate the structure change shown in Fig. \ref{fig1}. This
model space is, unfortunately, too large to perform a shell model
calculation. The model space that we can presently handle is a
truncated one $f_{5/2}p_{1/2}g_{9/2}d_{5/2}$. (Note that
calculation for $^{76}$Sr can be performed by using the
extrapolation method \cite{Mizusaki}.) Setting nucleons in the
$p_{3/2}$ orbit inactive is a severe restriction since
correlations in the $fp$ shell are strong. Nevertheless, this can
be compromised by adjusting effective interactions. The choice of
the present model space seems to have grasped the basic physics,
as we discuss below.

\begin{figure}[t]
\includegraphics[width=8cm,height=6.2cm]{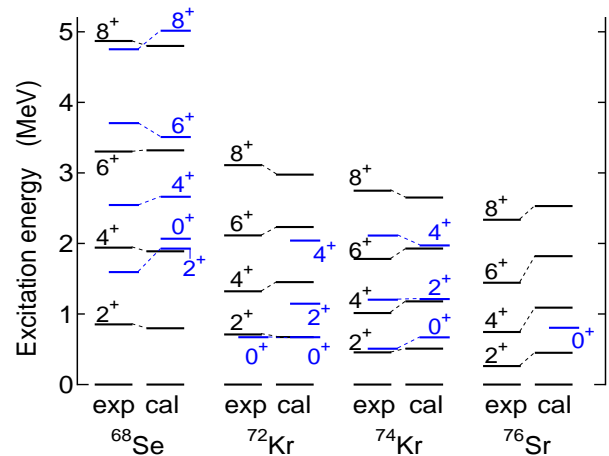}
\caption{(Color online) Comparison of experimental and calculated
level schemes for $^{68}$Se, $^{72}$Kr, $^{74}$Kr, and $^{76}$Sr.}
\label{fig2}
\end{figure}

With the extended $P+QQ$ Hamiltonian \cite{Kaneko,HaseN}, we have
searched for suitable parameters that can reproduce the
experimental data in this mass region. The level schemes shown in
Fig. \ref{fig2} are obtained by using the following set of
parameters. Single-particle energies for the $f_{5/2}$, $p_{1/2}$,
$g_{9/2}$, and $d_{5/2}$ orbits are 0.0, 0.3, 1.5 and 2.0 MeV,
respectively. For interaction strengths, we take
  \begin{eqnarray}
   & {} &  g_0 = 0.25(68/A), \quad g_2 = 0.16(68/A)^{5/3}, \nonumber \\
   & {} &  \chi_2 = 0.14(68/A)^{5/3},
         \chi_3 = 0.04(68/A)^2 \mbox{ in MeV}. \label{eq:2}
  \end{eqnarray}
The average $T=0$ monopole field and $T=1$ monopole corrections
(in MeV) are fixed as follows:
  \begin{eqnarray}
   & {} & k^0=H^{T=0}_{\rm mc}(a,b) = -0.63(68/A)
     \mbox{ for arbitrary } (a,b), \nonumber \\
   & {} & H^{T=1}_{\rm mc}(f_{5/2},p_{1/2}) = -0.6, \nonumber \\
   & {} & H^{T=1}_{\rm mc}(f_{5/2},g_{9/2}) =
     H^{T=1}_{\rm mc}(f_{5/2},d_{5/2}) = -0.5,  \nonumber \\
   & {} & H^{T=1}_{\rm mc}(g_{9/2},g_{9/2}) = 0.2,
    H^{T=1}_{\rm mc}(d_{5/2},d_{5/2}) = -0.18.
               \label{eq:3}
  \end{eqnarray}
It is found that to excite nucleons more efficiently to the upper
orbits $(g_{9/2},d_{5/2})$, we should take a small $d_{5/2}$
single-particle energy close to the $g_{9/2}$ one and large
monopole corrections $H^{T=1}_{\rm mc}(f_{5/2},g_{9/2})$ and
$H^{T=1}_{\rm mc} (f_{5/2},d_{5/2})$. The above parameter set may
be named ``the effective interaction of the small model".

Calculated level schemes for the $N=Z$ nuclei $^{68}$Se,
$^{72}$Kr, $^{76}$Sr, and the $N=Z+2$ nucleus $^{74}$Kr are
compared with experimental data in Fig. \ref{fig2}. Both the
ground band (in black) and the first excited $0^+$ band (in blue)
are shown for each nucleus. The truncated space without the
$p_{3/2}$ orbit cannot perfectly reproduce the data.  However, the
qualitative feature for each nucleus is well described. The
$^{68}$Se nucleus is seen to have a different structure with much
larger energy intervals than the other three nuclei. What we find
remarkable is that the calculation correctly gives the first
excited $0_2^+$ state in $^{72}$Kr below the $2_1^+$ one, in
agreement with the experimental finding \cite{Bouchez}.

To trace the microscopic origin of the structure changes, we
calculate nucleon occupation numbers $\langle n_a \rangle$ in
respective orbits. As we employ the isospin invariant Hamiltonian,
the proton and neutron occupation numbers in $N=Z$ nuclei are
equal to each other, {\it i.e.,} $\langle n_a^\pi \rangle =
\langle n_a^\nu \rangle$. To simplify the notation, we abbreviate
the lower orbits $(f_{5/2},p_{1/2})$ and the upper orbits
$(g_{9/2},d_{5/2})$ to ``$fp$" shell and ``$gd$" shell,
respectively. We suppose that the former represents the $fp$ shell
and the latter the $gd$ shell.


\begin{figure*}
\includegraphics[width=5cm,height=3.4cm]{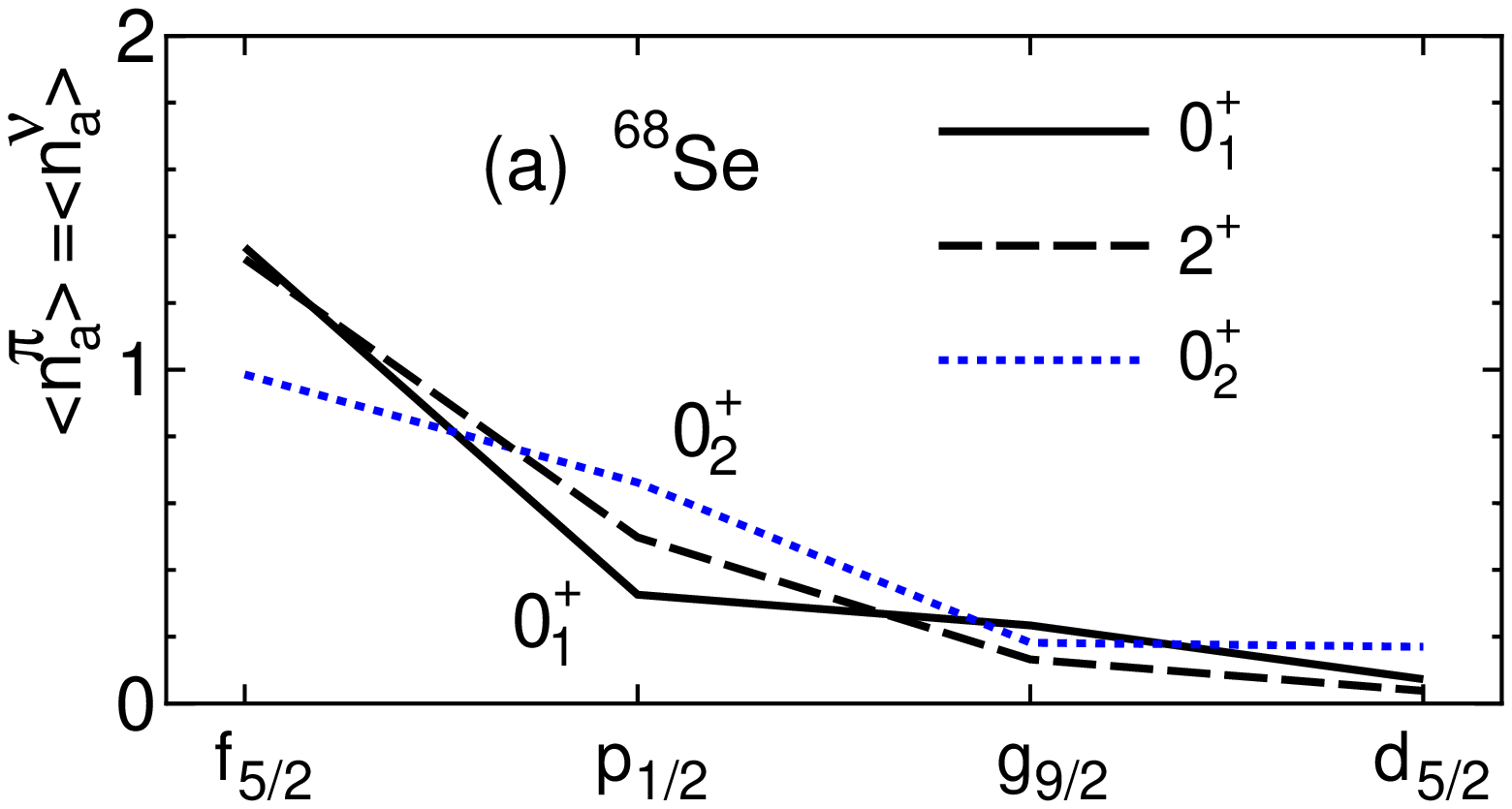}
\includegraphics[width=5cm,height=5cm]{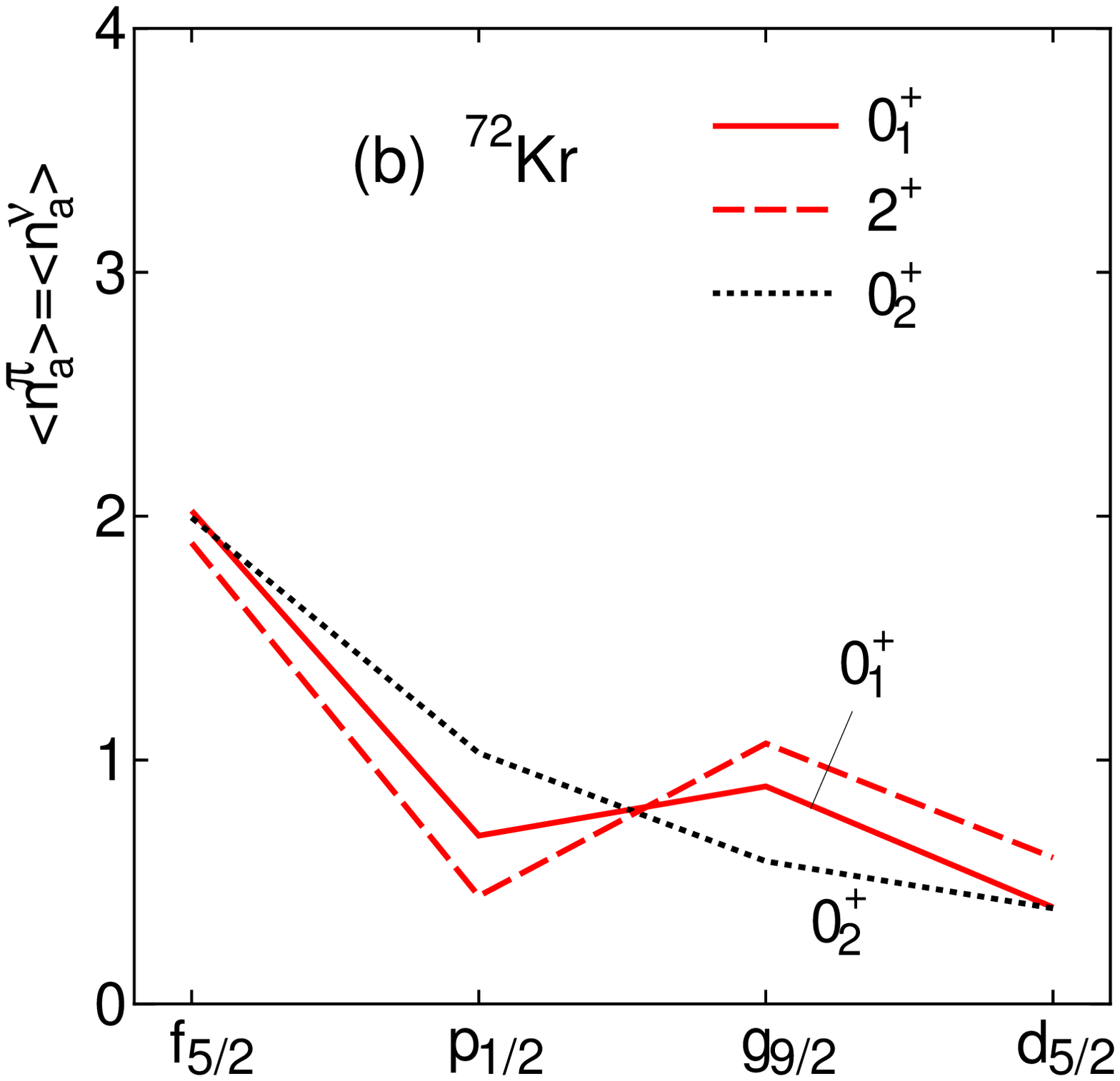}
\includegraphics[width=5cm,height=5cm]{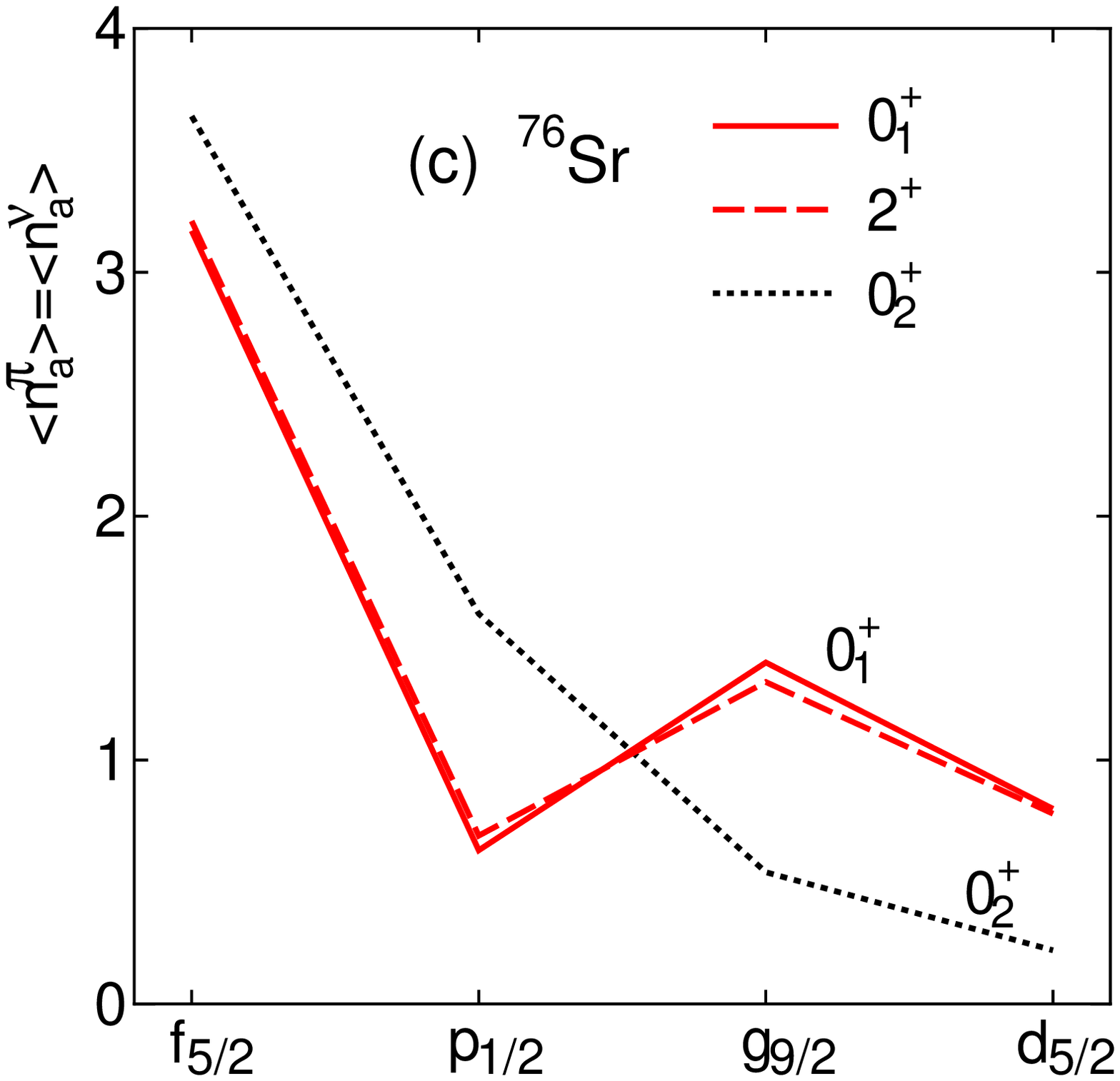}
  \caption{(Color online) Proton and neutron occupation numbers
          in $^{68}$Se, $^{72}$Kr, and $^{76}$Sr.}
  \label{fig3}
\end{figure*}

The calculated occupation numbers are shown in Fig. 3. As seen in
Fig. 3a, nucleons in $^{68}$Se have a negligible occupation in the
``$gd$" shell. Although the $0_2^+$ state has a different
configuration from the states of the ground band, all of the
states are basically constructed in the ``$fp$" shell. We should
mention that the description for $^{68}$Se with the present model
space may be too simplified. However, a more realistic calculation
for $^{68}$Se in the $pf_{5/2}g_{9/2}$ shell (including the
$p_{3/2}$ orbit) \cite{Kaneko} does not give a structure much
different from the present one. We have confirmed this by
improving the $pf_{5/2}g_{9/2}$ shell model so as to reproduce
well the observed coexisting oblate and prolate bands
\cite{Fischer1}. We thus conclude with confidence that the
$^{68}$Se nucleus does not essentially occupy the ``$gd$" shell.

In contrast to $^{68}$Se, approximately two protons and two
neutrons jump into the $``gd"$ shell in the ground state $0_1^+$
and the first excited state $2_1^+$ in $^{76}$Sr (see Fig. 3c).
Similarly, nucleons in $^{72}$Kr also start to occupy the $``gd"$
shell (see Fig. 3b). These results indicate that for $^{76}$Sr and
$^{72}$Kr, the $0_2^+$ state instead of the ground state has the
ordinary type of configuration, {\it i.e.,} the occupation number
deceases gradually as the single-particle energy increases. The
ground state has the dominant configuration
$(``fp")^{n-4}(``gd")^4$, the probability of which is about 0.45
in $^{72}$Kr and is more than 0.7 in $^{76}$Sr.

The wave functions of the $0_1^+$ and $0_2^+$ states in $^{72}$Kr
have the following weights for the leading configurations:
\begin{eqnarray}
  |0_1^+ \rangle & : & 0.22 (``fp")^8 + 0.45 (``fp")^4(``gd")^4
      + \cdots , \nonumber \\
  |0_2^+ \rangle & : & 0.47 (``fp")^8 + 0.40 (``fp")^4(``gd")^4
      + \cdots .    \label{eq:4}
\end{eqnarray}
These numbers show that the ground state is constructed starting
from the excited configuration $(``fp")^4(``gd")^4$ and the second
$0^+$ state is constructed starting from $(``fp")^8$. The two
$0^+$ states must reverse their order in energy if interactions
were increased gradually in a treatment of the perturbation
theory. This reversal manifests itself in the occupation numbers
in Fig. 3b; the $0_1^+$ state has less nucleons in the $``fp"$
shell and more nucleons in the $``gd"$ shell as compared to the
$0_2^+$ state. As shown in Fig. 3c, $^{76}$Sr has the same
characteristic occupation numbers obtained with the extrapolation
method.

The calculation thus reveals a large difference in occupation
numbers between $^{68}$Se and $^{72}$Kr, which correlates with
their qualitatively different picture in the $J-\omega$ graph in
Fig. \ref{fig1}. It has been known that states corresponding to
both oblate and prolate shapes coexist in $^{68}$Se
\cite{Fischer1}. Now our result that the ground state is
constructed mainly in the $fp$ shell suggests that the mean field
of $^{68}$Se is deformed but does not lie very far from the
spherical one of shell model. In contrast, the $^{72}$Kr ground
state cannot be described in the perturbation theory based on the
$``fp"$ configuration. This means that a decisive, sudden breaking
of the spherical mean field of shell model occurs when going from
$^{68}$Se to $^{72}$Kr. We may call the abrupt change in mean
field ``phase transition". Then what mean field is formed after
the breaking of the spherical shell model one?  The occupation of
the ``$gd$" shell in $^{72}$Kr and $^{76}$Sr implies a definite
formation of the Nilsson orbits in the deformed mean field. The
transition to $^{72}$Kr must be a shape phase transition. In fact,
the level scheme of $^{76}$Sr shows a rigid rotor behavior as seen
in Fig. \ref{fig1}, which is qualitatively different from the
nuclei with $N=Z\le 35$. It has been discussed with the deformed
mean field language \cite{Nazarewicz,Kobayasi} that there is a
rapid change in deformation with the deformation parameter $\beta$
from $\sim 0.25$ in $^{68}$Se, $\sim 0.4$ in $^{72}$Kr, and $\sim
0.5$ in $^{76}$Sr. The $g_{9/2}$ orbit has been considered to
drive deformation in this mass region \cite{Gade}.

It has been found that in $^{72}$Kr, the nearby $0_1^+$ and
$0_2^+$ states coexist with different shapes; the excited one is
considered as a shape isomer \cite{Bouchez,Sun2}. In our wave
functions (\ref{eq:4}), the $0_2^+$ state has a very large
component of the $(``fp")^4(``gd")^4$ configuration which is the
main component of the $0_1^+$ state, while the $0_1^+$ state has a
large component of the $(``fp")^8$ configuration which is the main
component of the $0_2^+$ state. The situation that these two $0^+$
states mutually have both configurations with considerable amount
in the wave functions suggests a rather strong coupling between
the two states. Such a strong coupling can explain why the moment
of inertia of the $2_1^+$ state in $^{72}$Kr deviates from the
regular rotor behavior (see Fig. 1); it is simply because the
$0_1^+$ energy is pushed down by the coupling.


\begin{figure}[h]
\includegraphics[width=7cm,height=6.8cm]{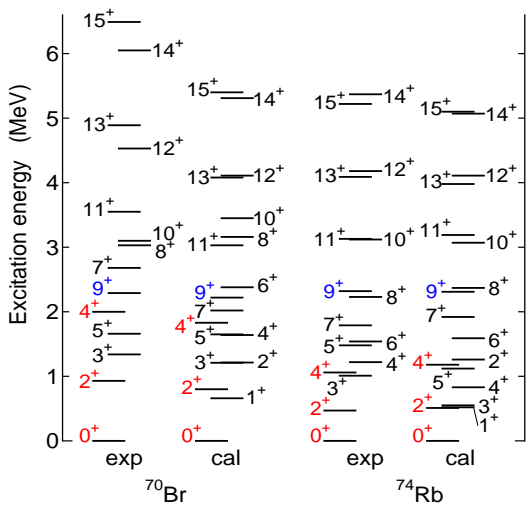}
  \caption{(Color online) Experimental and calculated level
           schemes for $^{70}$Br and $^{74}$Rb.}
  \label{fig4}
\end{figure}

To test the model's reliability, we should also calculate the
odd-odd nuclei $^{70}$Br and $^{74}$Rb. Calculated energy levels
are compared with experimental data \cite{Jenkins,Oleary} in Fig.
\ref{fig4}. The agreement between calculation and experiment is
satisfactory for the restricted model. It is found that the
configuration $(``fp")^6$ is dominant in the ground state of
$^{70}$Br. Contrary to this, for $^{74}$Rb the configuration
$(``fp")^6(``gd")^4$ is leading in the low-lying collective states
in the $T=1$ band ($0^+$,$2^+$,$4^+$) and $T=0$ band
($3^+$,$5^+$,$7^+$). These results are consistent with the clear
difference in their moments of inertia between $^{70}$Br and
$^{74}$Rb (Fig. \ref{fig1}). We thus reach the same conclusion
that the structure of $^{74}$Rb is qualitatively different from
that of $^{70}$Br and a shape phase transition takes place at
$Z=N=35$.

\begin{figure}[h]
\includegraphics[width=5.4cm,height=3.3cm]{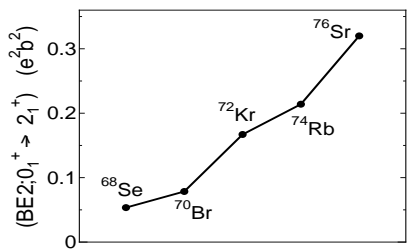}
  \caption{Calculated $B(E2)\hspace{-1.5mm}\uparrow$ values
           for $A=68-76$ nuclei.}
  \label{fig5}
\end{figure}

The structure change along the Ge isotopic chain manifests itself
also in a rapid increase of $B(E2;0_1^+ \rightarrow 2_1^+)$
\cite{HaseN}. Similar increase of $B(E2)\hspace{-1.5mm}\uparrow$
is expected when $N$ exceeds 35 along the $N=Z$ line. In Fig.
\ref{fig5} the calculated $B(E2)\hspace{-1.5mm}\uparrow$ values
from $^{68}$Se to $^{76}$Sr clearly show such a feature. Here, we
used the effective charges $e_\pi=1.5e$ and $e_\nu=0.5e$. From the
above analysis, the rapid increase in
$B(E2)\hspace{-1.5mm}\uparrow$ is attributed to the fact that more
degrees of freedom are opened up for the quadrupole correlation as
nucleons occupy the $``gd"$ shell. Thus the present results
support our previous prediction \cite{HaseN} that the notable
increase of $B(E2)\hspace{-1.5mm}\uparrow$ at $N=40$ in Ge
isotopes is caused by a jump of nucleons into the $``gd"$ shell.

We note that our truncated model space exposes its weakness in
reproduction of absolute value of $B(E2)\hspace{-1.5mm}\uparrow$.
For $^{72}$Kr, the calculated value 0.17 $e^2b^2$ is only about
one third of the observed one $\sim 0.5$ $e^2b^2$ \cite{Gade}.
However, we know the possible reason for the discrepancy. The
shell model for $^{68}$Se with the $pf_{5/2}g_{9/2}$ space
\cite{Kaneko} gave the $B(E2)\hspace{-1.5mm}\uparrow$ value 0.16
$e^2b^2$, which is about three times as large as the present
result 0.054 $e^2b^2$. Including the $p_{3/2}$ orbit, which
contributes to correlations in the $fp$ shell, will enhance the
$B(E2)\hspace{-1.5mm}\uparrow$ values considerably. For $^{68}$Se,
while the $pf_{5/2}g_{9/2}$ shell model \cite{Kaneko} reproduced a
positive spectroscopic quadrupole moment ($Q_{sp}$) corresponding
to an oblate shape for the $2_1^+$ state, the present model fails.
This suggests that the lack of the $p_{3/2}$ orbit is a serious
problem for those quantities sensitive to the details of wave
functions. Obviously it requires a larger model space to describe
these quantities.


In conclusion, the experimental level schemes of the $A \sim 70$
$N=Z$ nuclei show clearly an abrupt structure change at $N=Z=36$.
We have investigated this problem by means of large-scale shell
model calculations. In spite of the limit of the practicable model
space, we have been able to reproduce the basic features of the
level schemes. Through analyzing the wave functions, we have
concluded that the qualitative structure difference between the
nuclei with $N=Z\le 35$ and those with $N=Z\ge 36$ has the origin
of the upper $``gd"$ shell occupation. It has been discussed that
the structure change is caused by a decisive breaking of the
spherical shell model mean field and a formation of deformed mean
field. In this sense, we have witnessed a basic mechanism of the
phase transition to deformation in nuclei. We note that
calculations in a space larger than $pf_{5/2}g_{9/2}d_{5/2}$ must
be carried out to describe deformation properly.



\end{document}